\documentclass[sigconf]{acmart}

\usepackage{booktabs} 
\usepackage{tabularx}
\usepackage{tablefootnote}
\usepackage{todonotes}
\usepackage{multirow} 
\usepackage{amsmath}
\usepackage{changes} 
\definechangesauthor[name={Fabio}, color=blue]{fc}
\definechangesauthor[name={Mika}, color=brown]{mm}
\definechangesauthor[name={Maelick}, color=purple]{mc}

\setcopyright{rightsretained}

\acmDOI{10.475/123_4}

\acmISBN{123-4567-24-567/08/06}

\acmConference[MSR'2018]{ACM MSR conference}{May 28--29, 2018}{Gothenburg, 
Sweden} 
\acmYear{2018}
\copyrightyear{2018}

\acmArticle{4}
\acmPrice{15.00}

\copyrightyear{2018}
\acmYear{2018}
\acmConference[MSR '18]{MSR '18: 15th International Conference on Mining Software Repositories }{May 28--29, 2018}{Gothenburg, Sweden}
\acmPrice{15.00}
\acmDOI{10.1145/3196398.3196444}
\acmISBN{978-1-4503-5716-6/18/05}

\begin{document}

\title{Natural Language or Not (NLoN) - A Package for Software Engineering Text Analysis Pipeline}

\author{Mika V.~M{\"a}ntyl{\"a}}
\orcid{1234-5678-9012}
\affiliation{%
  \department{M3S}
  \institution{University of Oulu}
  \country{Finland} 
}
\email{mika.mantyla@oulu.fi}

\author{Fabio Calefato}
\orcid{0000-0003-2654-1588}
\affiliation{%
  \department{Dipartimento Jonico}
  \institution{University of Bari}
  \country{Italy} 
}
\email{fabio.calefato@uniba.it}

\author{Maelick Claes}
\orcid{0000-0003-2259-3946}
\affiliation{%
  \department{M3S}
  \institution{University of Oulu}
  \country{Finland} 
}
\email{maelick.claes@oulu.fi}


\begin{abstract}
The use of natural language processing (NLP) is gaining popularity in software engineering. In order to correctly perform NLP, we must pre-process the textual information to separate natural language from other information, such as log messages, that are often part of the communication in software engineering. We present a simple approach for classifying whether some textual input is natural language or not. Although our NLoN package relies on only 11 language features and character tri-grams, we are able to achieve an area under the ROC curve performances between 0.976-0.987 on three different data sources, with Lasso regression from Glmnet as our learner and two human raters for providing ground truth. Cross-source prediction performance is lower and has more fluctuation with top ROC performances from 0.913 to 0.980. Compared with prior work, our approach offers similar performance but is considerably more lightweight, making it easier to apply in software engineering text mining pipelines. Our source code and data are provided as an R-package for further improvements.
\end{abstract}

%
%



\maketitle

\section{Introduction}
 
Mining software repositories community often needs to utilize methods originating from the Natural Language Processing (NLP) community. Sentiment Analysis is an NLP task that has received a 50-fold increase in the number of papers between 2005 and 2015~\cite{MANTYLA201816}. In software engineering it has gained recent attention on detecting developers' emotions~\cite{islam2016towards,Jongeling2017,mantyla2017bootstrapping,Gachechiladze2017Anger} as well as opinions about software products used in the field~\cite{martin2017survey,guzman2015retrieving,maalej2015bug}. Topic modeling is another NLP task that discovers topics  occurring in a set of documents. For example, in software engineering it has been used to improve test case selection for manual testing~\cite{hemmati2017prioritizing} and to detect error prone software components~\cite{chen2017topic}. What is common in these NLP tasks is that they may produce incorrect results unless pre-processing is able to distinguish natural language from other textual elements that are common in software engineering context, such as source code, system configuration, and stack traces. 

In software engineering literature, we can find prior works on this topic. Bacchelli et al.~\cite{bacchelli2012content} presented a system for classifying content of software development emails. Their approach uses several parser and machine learning techniques and achieves an F-score of 0.945 for classifying whether a line is natural language(NL) or not, further classifying non-NL as junk, patch, or stack trace. Their tool seemed ideal for our problem, however, we were unable to either get it running or just extract their benchmark data from the database dump to a CSV file, due to complex database design.
Overall, their solution offered very good performances, but their design seemed excessive. Thus, we searched for other solutions. In earlier work, Bacchelli et al.~\cite{bacchelli2010extracting} used a simpler approach which was based on regular expressions alone and achieved F-scores as high as 0.89 in separating source code from email. As our task is to separate natural language from various forms of SE communication, this only partially matched our needs. Yet, we found this approach appealing due its simplicity and relatively good performance. Before the work by Bacchelli et al., Betternburg et al.~\cite{bettenburg2008extracting} showed how regular expressions and island parsing can be used in a single project to extract patches, stack traces, source code, and enumerations from bug reports with an Accuracy between 97.0 and 100.0. However, they do not consider the extraction of natural language, thus, the task is a bit different from ours. Furthermore, Betternburg et al. did not report how easily transferable their approach is between projects, thus making us unsure if one could develop a similarly accurate regular expression for diverse data sources. 

When looking at the natural language processing literature, we realized that detecting natural language from computing outputs could be viewed as a special case of a language detection task, e.g. whether a piece of text is written in English or French. From the language detection literature, e.g.~\cite{jauhiainen2017evaluation}, we learned that different language features can be used, and it seemed that very good performances are achieved by extracting character n-grams between three to five characters.
\section{Methodology}


Using ideas originating from software engineering and language detection, we aim at creating a lightweight classifier tool that would take a line of text as input and predict whether it is \underline{N}atural \underline{L}anguage \underline{o}r \underline{N}ot (NLoN). The tool is currently available as an open-source R package on GitHub ~\cite{NLoN_Rpackage}.

We  want the cost of moving our tool between projects manageable in terms of effort and expertise. To control effort, we decided that it should require no more than 2,000 lines of text for training. According to our experience, manually labeling a single line takes about 3 seconds meaning that 2,000 lines can be labeled in 1 hour and 40 minutes. However, counting in some breaks and time needed to think about borderline cases, we think it is realistic to budget 4 hours for annotating project-specific data that can be incorporated in adaptations of the tool. In order to minimize expertise and effort, we decided to use machine learning rather than regular expressions. Although regular expression and parsers can offer excellent accuracy~\cite{bettenburg2008extracting}, they may require significant expertise when adapting them from one project to another. The adaption of regular expressions also requires investigation of the project under study. We think that effort is better spent in manual data labeling, whose results are then fed to a machine learner that decides whether a particular regular expression is a good predictor or not. 

We utilized 3 data sources with 2,000 samples each. First, we have Comments made on the Mozilla issue tracker for projects where most of the professional development happens, i.e., Firefox, Core, and Firefox OS. In total, the repository has over 6 million issue comments. Second, we have Chat entries retrieved from the public Slack archive of Kubernetes. In particular, we downloaded ~16K random entries from the \texttt{\#kubernetes-dev} channel, where we expected to find more representative examples of chat entries mixing natural language with code. Third, we have Email messages mined from one of the mailing list archives of the Apache Lucene project. Similarly, we downloaded the entire content (25K message) of the \texttt{lucene-dev} mailing list, where we expected to find more emails containing natural language content interleaved with code snippets.

The first and second authors performed independent labeling. We noticed that oversight errors in human labeling occurred between 1 and 2\% of the labels for both labelers. After these errors were fixed, the labelers agreed on 97 to 98\% of the lines. To keep our presentation of results tidy, we only use the labels of the second rater (= second author). The first rater (= first author) was responsible for the NLoN-package implementation, thus, it is  possible that his ratings are influenced by the feature engineering done for the machine learning model. Thus, the use of the second rater who was not involved in the tool implementation offers unbiased text labels. Still, we note that there is no meaningful difference in our classifier performance between raters.   

For machine learning, we implemented two approaches: feature engineering (FEng) and character tri-grams (C3gram). Feature engineering is inspired by the success of regular expressions of past work~\cite{bettenburg2008extracting,bacchelli2010extracting}. Yet, we do not interpret regular expressions as absolute rules where matching certain condition would classify the input as NL or non-NL. Rather we extract them as language features and feed the results to our machine learning algorithm, e.g. if the line ends with ``\{'', it is fed as 1, or as 0 otherwise. Additionally, feature engineering uses statistics of each line such as the ratio of non-letter characters and the ratio of capital letters. All feature engineering predictors are shown in Table~\ref{tab:feng}. Ten of our eleven features were created when working with our first data set, but to our surprise these features also performed very well with the two other data sets. In the end, we only found one extra feature that improved performance. However, we think that there is room for improvement in future works.

Character tri-grams were suggested by language detection literature, e.g. \cite{jauhiainen2017evaluation}. We were afraid that due to our small sample size (2,000 lines) and the limited amount of contents each line holds, that tri-grams would not perform very well. Language detection approaches offer good performance starting from 25 characters and top performance is reached at 60 characters \cite{jauhiainen2017evaluation}, requirements that are not met by many of our input lines. Also, the number of samples in the language detection can go up to millions \cite{jauhiainen2017evaluation}.

Glmnet implements a generalized linear model with a penalty. It was chosen as our machine learning tool due to its fast performance, robustness, and ability to handle large and sparse input space with multiple correlating inputs~\cite{friedman2010regularization}. Due to these features, penalized regressions are regarded as a recommended strategy for natural language processing tasks~\cite{gentzkow2017text}. Glmnet performs both variable selection and regularization, which prevent over-fitting by limiting the model complexity with a penalty term lambda. This ensures that we can test language features with high dimensionality, e.g. character tri-grams, without having to worry about feature selection or over-fitting. The ability to do feature selection as part of prediction has made Glmnet gain interest in the defect prediction community as well~\cite{osman2017automatic}. We use Glmnet for performing binomial logistic lasso regression and optimize its 10-fold cross-validation performance with respect to the area under the ROC curve (AUC). We report a performance at lambda min which gives the maximum mean AUC in cross validation. We repeated the cross validation 5 times and use the median performance in our results to counter the effects of non-balanced data partitioning.
\section{Results}
\subsection{Within-source prediction}
Table~\ref{tab:performance} shows the results of machine learning using Glmnet with 10-fold cross validation. In all cases, we can see that both AUC and F-scores are above 0.9, and in many cases above 0.95. Combining both feature engineering and character tri-grams always offers a superior performance but it also contains the highest number of variables. The F-scores are shown to make backward comparison to previous papers easier. As reported in Bacchelli et al. \cite{bacchelli2012content}, their work resulted in F-scores up to 0.945. Our F-scores are between 0.959 and 0.970. From past work we found no execution times and our very brief execution time tests show that our tool can classify 1,000, 10,000, and 100,000 lines in 0.3s, 2.6s. and 27s respectively with a personal computer using a single core and logical unit.

For the remainder of this paper, we will only report and discuss AUC measures. Unlike F-measure, AUC is threshold-independent, i.e., it does not depend on selecting a \emph{cutoff} value that the model uses to decide whether an instance is to be classified as either positive or negative. Threshold-dependent measures can lead to different and conflicting conclusions~\cite{Rahman2012threshold}. Besides, the three experimental datasets described earlier are imbalanced: the instances of the class NL outnumber those in the other class non-NL (see Table~\ref{tab:performance}). This problem, referred to as class imbalance~\cite{He2009imbalance}, can significantly compromise both the performance of the learning algorithms and the reliability of the assessment metrics. However, because it is threshold-independent, AUC is better at providing a reliable measure of the classification performance in presence of imbalanced datasets~\cite{Huang2005imbalance}.

\begin{table}[]
\centering
\caption{Classification performance}
\label{tab:performance}
\begin{tabular}{lllll}
                         &               & Comments & Chat     & Email     \\ \hline
NL lines           &               		& 64.8 \%       & 83.4 \% & 63.6 \% \\ \hline
\multirow{3}{*}{AUC} & FEng 			& 0.984          & 0.971    & 0.969    \\
                         & C3gram  		& 0.973          & 0.951    & 0.962    \\
                         & Both  	    & 0.987          & 0.976    & 0.981    \\ \hline
\multirow{3}{*}{F1}      & FEng 		& 0.957          & 0.957    & 0.938    \\ 
                         & C3gram  		& 0.918          & 0.935     & 0.918    \\
                         & Both         & 0.959          & 0.970    & 0.959   
\end{tabular}
\end{table}

\subsection{Feature Engineering (FEng)}
Table ~\ref{tab:norm} shows normalized feature engineering coefficients at optimal penalty (\emph{lambda.min}) while Table ~\ref{tab:feng} explains each coefficient. A positive sign of a coefficient means that the coefficient increases the probability of a line being natural language, while negative values decrease it. We can notice that most of the coefficients have the same sign in all three data sets, meaning they predict towards the same end result. The coefficients are normalized, meaning that their size indicates their importance. An empty cell indicates that predictor is not selected for the model. For example, we can see that the number of stop-words, i.e. very common English words such ``it'' and ``the'', strongly predicts natural language while the ratio of special characters predicts the opposite.

For stop words, we used a list included in MySQL database but we removed source code specific words, e.g. ``for'' and ``while''. We included the number of stop words twice with different ways of tokenizing character streams to words as we found that, depending on the data, a different tokenization was required. Coefficient values show that our decision was correct as both stop-word predictors are meaningful for all three data sets. 

\begin{table}[]
\centering
\caption{Normalized coefficients for model FEng}
\label{tab:norm}
\begin{tabular}{lrrr}
             & Comments   & Chat         & Email        \\ \hline
r.caps       & -0.40 & -0.11 & 0.14  \\
r.specials   & -2.04 & -0.41 & -1.13 \\
r.numbers    & -0.90 & -0.21 & .     \\
l.words      & -1.59 & -0.85 & -2.54 \\
n.sw         & 1.80  & 2.31  & 0.91  \\
n.sw2        & 0.59  & 0.29  & 0.30  \\
last.c.code  & -0.15 & 0.04  & -0.63 \\
c1-3.letters & 0.29  & 0.06  & 0.33  \\
last.c.nl    & 1.42  & 0.90  & 0.50  \\
n.emoticons  & 0.64  & 0.05  & .     \\
first.c.at   & .     & 1.58  & .    
\end{tabular}
\end{table}

\begin{table}[]
\centering
\caption{Feature engineering predictors}
\label{tab:feng}
\begin{tabular}{ll}
Abbreviation   & Explanation \\ \hline
r.caps         & Ratio of capital letters                                      \\
r.specials     & Ratio of chars not alphanumeric or whitespace \\
r.numbers      & Ratio of numbers                                              \\
l.words        & Length of words                                               \\
n.sw           & Number of stop-words split with white space                    \\
n.sw2          & Number of stop-words split with tokenize\_words                 \\
last.c.code    & Is last character typical in code , e.g. \{ or ;      \\
n.c1-3.letters & Number of letters in first three characters                   \\
last.c.nl      & Is last character typical in NL, e.g. ? or .       \\
n.emoticons    & Number of emoticons                                           \\
first.c.at     & Is first character of line @-sign                            
\end{tabular}
\end{table}

\subsection{Character tri-grams (C3gram)}
For character trigrams, we did some pre-processing. We changed all numbers to zeros as we figured that recognizing numbers would be important but the exact numbers would not matter for our task. We also converted all characters to lower case as we noticed no performance difference when keeping casing. We do realize that ratio of capital letters is a predictor in feature engineering, but still, as keeping capitals offered no performance improvement, we removed them. Perhaps with larger training data it would be meaningful.

Table ~\ref{tab:3gram} shows the number of selected predictors and trigrams at minimum lambda, which gives the maximum AUC, but also for 1se lambda, which gives the most regularized (penalized) model such that AUC is within one standard deviation of the maximum. Utilizing not the best model but one that is one standard deviation away from it (lambda 1se) is an heuristic often used in machine learning, when several predictors are present, which slightly sacrifices model accuracy to select a simpler model whose accuracy is similar to the best model~\cite{friedman2001elements,krstajic2014cross}

We have many character trigrams in our input data, between 8740 and 11169, but in all cases less than 10\% of those are selected as predictors for the best model. The number of selected tri-grams varies between 304 and 597. When we go for the simpler model, whose performance is within one standard deviation from the best model, we find that for the Chat messages from Kubernetes only 1.4\% of the trigrams (149/11169) are used in prediction. In the case of the Comments from Mozilla and the Mails from Lucene, the percentages are 2.8 and 2.4 percent respectively.

The reduction in the number of predictors of the simpler model is even more evident in the model combining both tri-grams and feature engineering (see Table ~\ref{tab:both}). The best model, using both, has a number of predictors between 138 and 447, while the simpler model, giving nearly identical performance, uses between 27 to 68 predictors.

\begin{table}[]
\centering
\caption{Number of predictors and performance at two different lambda values for model C3gram}
\label{tab:3gram}
\begin{tabular}{lrrr}
                 & Comments & Chat  & Email  \\ \hline
C3grams        		& 11169    & 9101  & 8740  \\\hline
AUC (lambda min)      & 0.973    & 0.951 & 0.962 \\
selected 3-grams 	& 464      & 304   & 554   \\\hline
AUC (lambda 1se)      & 0.966    & 0.949 & 0.959 \\
selected 3-grams 	& 133      & 129   & 210  
\end{tabular}
\end{table}

\begin{table}[]
\centering
\caption{Number of predictors and performance at two different lambda values for model Both}
\label{tab:both}
\begin{tabular}{lrrr}
                 & Comments & Chat  & Email  \\ \hline
Features (C3grams+FEng)        & 11180    & 9112  & 8751  \\\hline
AUC (lambda min) 	  & 0.987    & 0.976 & 0.981 \\
selected features		& 406      & 138   & 383   \\\hline
AUC (lambda 1se)      & 0.986    & 0.973 & 0.980 \\
selected features	 & 48      & 27   & 68  
\end{tabular}
\end{table}

\subsection{Cross-source prediction}
We were also interested on how would our tool perform in cross-source prediction task. Table~\ref{tab:cross} shows these results. The first column (i) shows the results of the 10-fold cross validation using all the six thousand samples of the three source. We can see that in comparison to using just source-specific data (see Tables~\ref{tab:performance} and \ref{tab:cross}) the performance of character tri-grams slightly improves as it is higher than the midpoint of the range of within source prediction (0.968 vs. range 0.951-0.973), while the performance of feature engineering slightly reduces (0.970 vs range 0.969-0.984). This matches our expectation that character tri-grams do better with larger data sets as they are sparse in comparison to feature engineering numbers that can be computed from every line. When using both feature engineering and character tri-grams with all data, we achieve a performance of 0.982, while the performance of using only source specific data gives results varying between 0.987 and 0.976. We conclude that using all the data neither improves nor weakens the performance.

For cross-source prediction (see the last three columns in Table~\ref{tab:cross}), we can see that using the Kubernetes Chat messages and the Lucene Email messages to predict Mozilla issue Comments (ii) works out surprisingly well with AUC up to 0.980, which is almost as good as using Mozilla's own data (AUC 0.987, see Table ~\ref{tab:performance}). On the other hand using Mozilla issue Comments and Kubernetes Chat messages to predict Lucene Emails (iv)  offers much weaker performances with the best AUC at 0.913, which is considerable weaker than using Lucene mailing list's own data (0.981).

Our cross-prediction results show that directly using our data one can get very good performances when filtering out non-natural language text in a software engineering context. Nevertheless, we recommend labeling a data set for each source as the effort is quite low (estimated only four hours) and the performance is very likely to be better.  

\begin{table}[]
\centering
\caption{Cross-source prediction results (AUC)}
\label{tab:cross}
\begin{tabular}{lrrrr}
              & (i) & (ii) & (iii)  & (iv)  \\
              & All (CV) & Comments & Chat  & Email  \\ \hline
F-engineering & 0.970    & 0.975    & 0.964 & 0.911 \\
Char 3-grams  & 0.968    & 0.946    & 0.914 & 0.880 \\
Both          & 0.982    & 0.980    & 0.957 & 0.913
\end{tabular}
\end{table}

\subsection{Limitations and future work}
Our approach is relatively simple and offers very good performance in three different source types from three different projects. However, the results from three sources cannot be used to claim that our solution would work in all other software engineering contexts. In addition, we only tried one machine learning method (i.e., Glmnet) with default settings, and it is possible that other algorithms could offer better results.  Therefore, we welcome others to try and improve our solution that is available online alongside our data.

It is typical to have a mix of natural language and code in a line of text. When labeling we always consider these mixed lines as natural language lines. Based on reviewer feedback, we think it might be worthwhile to have a third class for the mixed lines. Such mixed lines would require further processing and they would need to be fed to another tool implementing parsing to separate the contents. Overall, one could challenge our choice of using line granularity, which was selected as we aimed for simplicity rather than perfection. Furthermore, we note that we did not assessed how NLoN performs with respect to different programming languages. 

Finally, the tool is only tested with an English language context and uses English stop-word list as part of the detection. The minimal requirement to use this tool in another language context would be to replace the English stop-word list with one of the corresponding language. Languages with numerous conjugated forms would probably also need lemmatization and pre-processing before our tool could be used.

\section{Conclusions}

In this paper, we have presented a solution to separate natural language from other text inputs that are common in software engineering such as stack traces or source code. From the software engineering domain, we derived the idea of using regular expressions to separate input in different types. However, we do not follow regular expression matches as absolute rules but rather as information that is fed to machine learning. We also extract other language features such as the ratio of capital letters and the number of the most common English words, i.e. stop words. Finally, from the language detection literature, we borrowed the idea of extracting character tri-grams as further information to feed our machine learning model. Our best model achieves an area under ROC curve performance from 0.976 to 0.987 in three different source types (bug tracker issue comments, chat messages, and email) which originate from three different projects (Mozilla Firefox, Kubernetes and Apache Lucene).  

When we originally came up with the problem that natural language should be separated from non-natural language while performing NLP tasks like as sentiment analysis or topic modeling, we were sure that a solution to this problem would be openly available. We found only one solution \cite{bacchelli2012content} that address the same problem and was openly available. Unfortunately, (i) we could not run it; (ii) even if we could have, the complexity of the solution seemed too high given the problem. Therefore, we implemented a lightweight solution of few hundreds lines of R-code and data files, instead of database dumps. Our solution is available as an open source R-package on GitHub ~\cite{NLoN_Rpackage}.

\bibliographystyle{ACM-Reference-Format}
\bibliography{99_refs_Bib} 


\begin{thebibliography}{00}


\ifx \showCODEN    \undefined \def \showCODEN     #1{\unskip}     \fi
\ifx \showDOI      \undefined \def \showDOI       #1{{\tt DOI:}\penalty0{#1}\ }
  \fi
\ifx \showISBNx    \undefined \def \showISBNx     #1{\unskip}     \fi
\ifx \showISBNxiii \undefined \def \showISBNxiii  #1{\unskip}     \fi
\ifx \showISSN     \undefined \def \showISSN      #1{\unskip}     \fi
\ifx \showLCCN     \undefined \def \showLCCN      #1{\unskip}     \fi
\ifx \shownote     \undefined \def \shownote      #1{#1}          \fi
\ifx \showarticletitle \undefined \def \showarticletitle #1{#1}   \fi
\ifx \showURL      \undefined \def \showURL       #1{#1}          \fi
\providecommand\bibfield[2]{#2}
\providecommand\bibinfo[2]{#2}
\providecommand\natexlab[1]{#1}

\bibitem[\protect\citeauthoryear{Bacchelli, Dal~Sasso, D'Ambros, and
  Lanza}{Bacchelli et~al\mbox{.}}{2012}]%
        {bacchelli2012content}
\bibfield{author}{\bibinfo{person}{Alberto Bacchelli}, \bibinfo{person}{Tommaso
  Dal~Sasso}, \bibinfo{person}{Marco D'Ambros}, {and} \bibinfo{person}{Michele
  Lanza}.} \bibinfo{year}{2012}\natexlab{}.
\newblock \showarticletitle{Content classification of development emails}. In
  \bibinfo{booktitle}{{\em Software Engineering (ICSE), 2012 34th Int'l Conf.
  on}}. IEEE, \bibinfo{pages}{375--385}.
\newblock


\bibitem[\protect\citeauthoryear{Bacchelli, D'Ambros, and Lanza}{Bacchelli
  et~al\mbox{.}}{2010}]%
        {bacchelli2010extracting}
\bibfield{author}{\bibinfo{person}{Alberto Bacchelli}, \bibinfo{person}{Marco
  D'Ambros}, {and} \bibinfo{person}{Michele Lanza}.}
  \bibinfo{year}{2010}\natexlab{}.
\newblock \showarticletitle{Extracting source code from e-mails}. In
  \bibinfo{booktitle}{{\em Program Comprehension (ICPC), 2010 IEEE 18th Int'l
  Conf. on}}. IEEE, \bibinfo{pages}{24--33}.
\newblock


\bibitem[\protect\citeauthoryear{Bettenburg, Premraj, Zimmermann, and
  Kim}{Bettenburg et~al\mbox{.}}{2008}]%
        {bettenburg2008extracting}
\bibfield{author}{\bibinfo{person}{Nicolas Bettenburg}, \bibinfo{person}{Rahul
  Premraj}, \bibinfo{person}{Thomas Zimmermann}, {and} \bibinfo{person}{Sunghun
  Kim}.} \bibinfo{year}{2008}\natexlab{}.
\newblock \showarticletitle{Extracting structural information from bug
  reports}. In \bibinfo{booktitle}{{\em Proc. of 2008 Int'l Working Conf. on
  Mining Software Repositories}}. ACM, \bibinfo{pages}{27--30}.
\newblock


\bibitem[\protect\citeauthoryear{Chen, Shang, Nagappan, Hassan, and
  Thomas}{Chen et~al\mbox{.}}{2017}]%
        {chen2017topic}
\bibfield{author}{\bibinfo{person}{Tse-Hsun Chen}, \bibinfo{person}{Weiyi
  Shang}, \bibinfo{person}{Meiyappan Nagappan}, \bibinfo{person}{Ahmed~E
  Hassan}, {and} \bibinfo{person}{Stephen~W Thomas}.}
  \bibinfo{year}{2017}\natexlab{}.
\newblock \showarticletitle{Topic-based software defect explanation}.
\newblock \bibinfo{journal}{{\em Journal of Systems and Software\/}}
  \bibinfo{volume}{129} (\bibinfo{year}{2017}), \bibinfo{pages}{79--106}.
\newblock


\bibitem[\protect\citeauthoryear{Friedman, Hastie, and Tibshirani}{Friedman
  et~al\mbox{.}}{2001}]%
        {friedman2001elements}
\bibfield{author}{\bibinfo{person}{Jerome Friedman}, \bibinfo{person}{Trevor
  Hastie}, {and} \bibinfo{person}{Robert Tibshirani}.}
  \bibinfo{year}{2001}\natexlab{}.
\newblock \bibinfo{booktitle}{{\em The elements of statistical learning}}.
  \bibinfo{volume}{Vol.~1}.
\newblock Springer series in statistics New York.
\newblock


\bibitem[\protect\citeauthoryear{Friedman, Hastie, and Tibshirani}{Friedman
  et~al\mbox{.}}{2010}]%
        {friedman2010regularization}
\bibfield{author}{\bibinfo{person}{Jerome Friedman}, \bibinfo{person}{Trevor
  Hastie}, {and} \bibinfo{person}{Rob Tibshirani}.}
  \bibinfo{year}{2010}\natexlab{}.
\newblock \showarticletitle{Regularization paths for generalized linear models
  via coordinate descent}.
\newblock \bibinfo{journal}{{\em J. of statistical software\/}}
  \bibinfo{volume}{{33}, 1} (\bibinfo{year}{2010}), \bibinfo{pages}{1}.
\newblock


\bibitem[\protect\citeauthoryear{Gachechiladze, Lanubile, Novielli, and
  Serebrenik}{Gachechiladze et~al\mbox{.}}{2017}]%
        {Gachechiladze2017Anger}
\bibfield{author}{\bibinfo{person}{Daviti Gachechiladze},
  \bibinfo{person}{Filippo Lanubile}, \bibinfo{person}{Nicole Novielli}, {and}
  \bibinfo{person}{Alexander Serebrenik}.} \bibinfo{year}{2017}\natexlab{}.
\newblock \showarticletitle{Anger and Its Direction in Collaborative Software
  Development}. In \bibinfo{booktitle}{{\em Proc. of Int'l Conf. on Software
  Engineering: New Ideas and Emerging Results Track}} \bibinfo{series}{{\em
  (ICSE-NIER '17)}}. IEEE Press, Piscataway, NJ, USA, \bibinfo{pages}{11--14}.
\newblock
\showISBNx{978-1-5386-2675-7}
\showDOI{%
\url{http://dx.doi.org/10.1109/ICSE-NIER.2017.18}}


\bibitem[\protect\citeauthoryear{Gentzkow, Kelly, and Taddy}{Gentzkow
  et~al\mbox{.}}{2017}]%
        {gentzkow2017text}
\bibfield{author}{\bibinfo{person}{Matthew Gentzkow}, \bibinfo{person}{Bryan~T
  Kelly}, {and} \bibinfo{person}{Matt Taddy}.} \bibinfo{year}{2017}\natexlab{}.
\newblock \bibinfo{booktitle}{{\em Text as data}}.
\newblock {T}echnical {R}eport. National Bureau of Economic Research.
\newblock


\bibitem[\protect\citeauthoryear{Guzman, Aly, and Bruegge}{Guzman
  et~al\mbox{.}}{2015}]%
        {guzman2015retrieving}
\bibfield{author}{\bibinfo{person}{Emitza Guzman}, \bibinfo{person}{Omar Aly},
  {and} \bibinfo{person}{Bernd Bruegge}.} \bibinfo{year}{2015}\natexlab{}.
\newblock \showarticletitle{Retrieving diverse opinions from app reviews}. In
  \bibinfo{booktitle}{{\em Empirical Software Engineering and Measurement
  (ESEM), 2015 ACM/IEEE Int'l Symposium on}}. IEEE, \bibinfo{pages}{1--10}.
\newblock


\bibitem[\protect\citeauthoryear{He and Garcia}{He and Garcia}{2009}]%
        {He2009imbalance}
\bibfield{author}{\bibinfo{person}{Haibo He} {and} \bibinfo{person}{Edwardo~A.
  Garcia}.} \bibinfo{year}{2009}\natexlab{}.
\newblock \showarticletitle{Learning from Imbalanced Data}.
\newblock \bibinfo{journal}{{\em IEEE Trans. on Knowl. and Data Eng.\/}}
  \bibinfo{volume}{{21}, 9} (\bibinfo{date}{Sept.} \bibinfo{year}{2009}),
  \bibinfo{pages}{1263--1284}.
\newblock
\showISSN{1041-4347}
\showDOI{%
\url{http://dx.doi.org/10.1109/TKDE.2008.239}}


\bibitem[\protect\citeauthoryear{Hemmati, Fang, M{\"a}ntyl{\"a}, and
  Adams}{Hemmati et~al\mbox{.}}{2017}]%
        {hemmati2017prioritizing}
\bibfield{author}{\bibinfo{person}{Hadi Hemmati}, \bibinfo{person}{Zhihan
  Fang}, \bibinfo{person}{Mika~V M{\"a}ntyl{\"a}}, {and} \bibinfo{person}{Bram
  Adams}.} \bibinfo{year}{2017}\natexlab{}.
\newblock \showarticletitle{Prioritizing manual test cases in rapid release
  environments}.
\newblock \bibinfo{journal}{{\em Software Testing, Verification and
  Reliability\/}} \bibinfo{volume}{{27}, 6} (\bibinfo{year}{2017}).
\newblock


\bibitem[\protect\citeauthoryear{Huang and Ling}{Huang and Ling}{2005}]%
        {Huang2005imbalance}
\bibfield{author}{\bibinfo{person}{Jin Huang} {and} \bibinfo{person}{Charles~X.
  Ling}.} \bibinfo{year}{2005}\natexlab{}.
\newblock \showarticletitle{Using AUC and Accuracy in Evaluating Learning
  Algorithms}.
\newblock \bibinfo{journal}{{\em IEEE Trans. on Knowl. and Data Eng.\/}}
  \bibinfo{volume}{{17}, 3} (\bibinfo{date}{March} \bibinfo{year}{2005}),
  \bibinfo{pages}{299--310}.
\newblock
\showISSN{1041-4347}
\showDOI{%
\url{http://dx.doi.org/10.1109/TKDE.2005.50}}


\bibitem[\protect\citeauthoryear{Islam and Zibran}{Islam and Zibran}{2016}]%
        {islam2016towards}
\bibfield{author}{\bibinfo{person}{Md~Rakibul Islam} {and}
  \bibinfo{person}{Minhaz~F Zibran}.} \bibinfo{year}{2016}\natexlab{}.
\newblock \showarticletitle{Towards understanding and exploiting developers'
  emotional variations in software engineering}. In \bibinfo{booktitle}{{\em
  Software Engineering Research, Management and Applications (SERA), 2016 IEEE
  14th Int'l Conf. on}}. IEEE, IEEE, \bibinfo{pages}{185--192}.
\newblock


\bibitem[\protect\citeauthoryear{Jauhiainen, Lind{\'e}n, and
  Jauhiainen}{Jauhiainen et~al\mbox{.}}{2017}]%
        {jauhiainen2017evaluation}
\bibfield{author}{\bibinfo{person}{Tommi Jauhiainen}, \bibinfo{person}{Krister
  Lind{\'e}n}, {and} \bibinfo{person}{Heidi Jauhiainen}.}
  \bibinfo{year}{2017}\natexlab{}.
\newblock \showarticletitle{Evaluation of language identification methods using
  285 languages}. In \bibinfo{booktitle}{{\em Proceedings of the 21st Nordic
  Conf. on Computational Linguistics, NoDaLiDa, 22-24 May 2017, Gothenburg,
  Sweden}}. Link{\"o}ping University Electronic Press,
  \bibinfo{pages}{183--191}.
\newblock


\bibitem[\protect\citeauthoryear{Jongeling, Sarkar, Datta, and
  Serebrenik}{Jongeling et~al\mbox{.}}{2017}]%
        {Jongeling2017}
\bibfield{author}{\bibinfo{person}{Robbert Jongeling},
  \bibinfo{person}{Proshanta Sarkar}, \bibinfo{person}{Subhajit Datta}, {and}
  \bibinfo{person}{Alexander Serebrenik}.} \bibinfo{year}{2017}\natexlab{}.
\newblock \showarticletitle{On negative results when using sentiment analysis
  tools for software engineering research}.
\newblock \bibinfo{journal}{{\em Empirical Software Engineering\/}}
  \bibinfo{volume}{{22}, 5} (\bibinfo{date}{01 Oct} \bibinfo{year}{2017}),
  \bibinfo{pages}{2543--2584}.
\newblock
\showISSN{1573-7616}
\showDOI{%
\url{http://dx.doi.org/10.1007/s10664-016-9493-x}}


\bibitem[\protect\citeauthoryear{Krstajic, Buturovic, Leahy, and
  Thomas}{Krstajic et~al\mbox{.}}{2014}]%
        {krstajic2014cross}
\bibfield{author}{\bibinfo{person}{Damjan Krstajic},
  \bibinfo{person}{Ljubomir~J Buturovic}, \bibinfo{person}{David~E Leahy},
  {and} \bibinfo{person}{Simon Thomas}.} \bibinfo{year}{2014}\natexlab{}.
\newblock \showarticletitle{Cross-validation pitfalls when selecting and
  assessing regression and classification models}.
\newblock \bibinfo{journal}{{\em Journal of cheminformatics\/}}
  \bibinfo{volume}{{6}, 1} (\bibinfo{year}{2014}), \bibinfo{pages}{10}.
\newblock


\bibitem[\protect\citeauthoryear{Maalej and Nabil}{Maalej and Nabil}{2015}]%
        {maalej2015bug}
\bibfield{author}{\bibinfo{person}{Walid Maalej} {and} \bibinfo{person}{Hadeer
  Nabil}.} \bibinfo{year}{2015}\natexlab{}.
\newblock \showarticletitle{Bug report, feature request, or simply praise? on
  automatically classifying app reviews}. In \bibinfo{booktitle}{{\em
  Requirements Engineering Conf. (RE), 2015 IEEE 23rd Int'l}}. IEEE,
  \bibinfo{pages}{116--125}.
\newblock


\bibitem[\protect\citeauthoryear{M\"{a}ntyl\"{a}, Calefato, and
  Claes}{M\"{a}ntyl\"{a} et~al\mbox{.}}{2018}]%
        {NLoN_Rpackage}
\bibfield{author}{\bibinfo{person}{Mika~V. M\"{a}ntyl\"{a}},
  \bibinfo{person}{Fabio Calefato}, {and} \bibinfo{person}{Ma\"{e}lick Claes}.}
  \bibinfo{year}{2018}\natexlab{}.
\newblock \bibinfo{booktitle}{{\em NLoN -- Natural Language or Not}}.
\newblock
\showURL{%
\url{https://github.com/M3SOulu/NLoN}}
\newblock
\shownote{R package version 0.1.0.}


\bibitem[\protect\citeauthoryear{Mantyla, Graziotin, and Kuutila}{Mantyla
  et~al\mbox{.}}{2018}]%
        {MANTYLA201816}
\bibfield{author}{\bibinfo{person}{Mika~V. Mantyla}, \bibinfo{person}{Daniel
  Graziotin}, {and} \bibinfo{person}{Miikka Kuutila}.}
  \bibinfo{year}{2018}\natexlab{}.
\newblock \showarticletitle{The evolution of sentiment analysis -- A review of
  research topics, venues, and top cited papers}.
\newblock \bibinfo{journal}{{\em Computer Science Review\/}}
  \bibinfo{volume}{27} (\bibinfo{year}{2018}), \bibinfo{pages}{16 -- 32}.
\newblock
\showISSN{1574-0137}
\showDOI{%
\url{http://dx.doi.org/https://doi.org/10.1016/j.cosrev.2017.10.002}}


\bibitem[\protect\citeauthoryear{M{\"a}ntyl{\"a}, Novielli, Lanubile, Claes,
  and Kuutila}{M{\"a}ntyl{\"a} et~al\mbox{.}}{2017}]%
        {mantyla2017bootstrapping}
\bibfield{author}{\bibinfo{person}{Mika~V M{\"a}ntyl{\"a}},
  \bibinfo{person}{Nicole Novielli}, \bibinfo{person}{Filippo Lanubile},
  \bibinfo{person}{Ma{\"e}lick Claes}, {and} \bibinfo{person}{Miikka Kuutila}.}
  \bibinfo{year}{2017}\natexlab{}.
\newblock \showarticletitle{Bootstrapping a lexicon for emotional arousal in
  software engineering}. In \bibinfo{booktitle}{{\em Proc. of Int'l Conf. on
  Mining Software Repositories}}. \bibinfo{pages}{198--202}.
\newblock


\bibitem[\protect\citeauthoryear{Martin, Sarro, Jia, Zhang, and Harman}{Martin
  et~al\mbox{.}}{2017}]%
        {martin2017survey}
\bibfield{author}{\bibinfo{person}{William Martin}, \bibinfo{person}{Federica
  Sarro}, \bibinfo{person}{Yue Jia}, \bibinfo{person}{Yuanyuan Zhang}, {and}
  \bibinfo{person}{Mark Harman}.} \bibinfo{year}{2017}\natexlab{}.
\newblock \showarticletitle{A survey of app store analysis for software
  engineering}.
\newblock \bibinfo{journal}{{\em IEEE Transactions on Software Engineering\/}}
  \bibinfo{volume}{{43}, 9} (\bibinfo{year}{2017}), \bibinfo{pages}{817--847}.
\newblock


\bibitem[\protect\citeauthoryear{Osman, Ghafari, and Nierstrasz}{Osman
  et~al\mbox{.}}{2017}]%
        {osman2017automatic}
\bibfield{author}{\bibinfo{person}{Haidar Osman}, \bibinfo{person}{Mohammad
  Ghafari}, {and} \bibinfo{person}{Oscar Nierstrasz}.}
  \bibinfo{year}{2017}\natexlab{}.
\newblock \showarticletitle{Automatic feature selection by regularization to
  improve bug prediction accuracy}. In \bibinfo{booktitle}{{\em Machine
  Learning Techniques for Software Quality Evaluation (MaLTeSQuE), IEEE
  Workshop on}}. IEEE, \bibinfo{pages}{27--32}.
\newblock


\bibitem[\protect\citeauthoryear{Rahman, Posnett, and Devanbu}{Rahman
  et~al\mbox{.}}{2012}]%
        {Rahman2012threshold}
\bibfield{author}{\bibinfo{person}{Foyzur Rahman}, \bibinfo{person}{Daryl
  Posnett}, {and} \bibinfo{person}{Premkumar Devanbu}.}
  \bibinfo{year}{2012}\natexlab{}.
\newblock \showarticletitle{Recalling the "Imprecision" of Cross-project Defect
  Prediction}. In \bibinfo{booktitle}{{\em Proc. of 20th Int'l Symposium on the
  Foundations of Software Engineering}} \bibinfo{series}{{\em (FSE '12)}}. ACM,
  New York, NY, USA, Article \bibinfo{articleno}{61}, 11 pages.
\newblock
\showISBNx{978-1-4503-1614-9}
\showDOI{%
\url{http://dx.doi.org/10.1145/2393596.2393669}}


\end{thebibliography}

\end{document}